\documentclass[runningheads]{llncs}

\usepackage[T1]{fontenc}
\usepackage{graphicx}
\usepackage[misc]{ifsym}
\usepackage{amsfonts}
\usepackage{amssymb}

\usepackage{amsthm}
\usepackage{amsmath}
\usepackage{mathrsfs}
\usepackage{indentfirst}
\usepackage{multirow}
\usepackage{bm}
\usepackage{color,xcolor}
\usepackage{footnote}
\usepackage{float}
\usepackage{lipsum}
\begin{document}
\title{A Localization-to-Segmentation Framework for Automatic Tumor Segmentation in Whole-Body PET/CT Images}
\titlerunning{L2SNet for Automatic Tumor Segmentation in PET/CT Images}
%
%
\author{Linghan Cai\inst{1}       \and
        Jianhao Huang\inst{2}     \and
        Zihang Zhu\inst{1}        \and
        Jinpeng Lu\inst{2}        \and \\
        Yongbing Zhang\inst{1}$^{(\textrm{\Letter})}$
}
\authorrunning{L. Cai, et al.}
%
\institute{School of Computer Science and Technology, Harbin Institute of Technology, Shenzhen, China  \\
\email{cailh@stu.hit.edu.cn, ybzhang08@hit.edu.cn}\\ \and
School of Science, Harbin Institute of Technology, Shenzhen, China\\}
\maketitle              

\renewcommand{\thefootnote}{}
\footnote{L. Cai and J. Huang contributed equally to this work.}

\begin{abstract}
Fluorodeoxyglucose (FDG) positron emission tomography (PET) combined with computed tomography (CT) is considered the primary solution for detecting some cancers, such as lung cancer and melanoma. Automatic segmentation of tumors in PET/CT images can help reduce doctors’ workload, thereby improving diagnostic quality. However, precise tumor segmentation is challenging due to the small size of many tumors and the similarity of high-uptake normal areas to the tumor regions. To address these issues, this paper proposes a localization-to-segmentation framework (L2SNet) for precise tumor segmentation. L2SNet first localizes the possible lesions in the lesion localization phase and then uses the location cue to shape the segmentation results in the lesion segmentation phase. To further improve the segmentation performance of L2SNet, we design an adaptive threshold scheme that takes the segmentation results of the two phases into consideration. The experiment with the MICCAI 2023 Automated Lesion Segmentation in Whole-Body FDG-PET/CT challenge dataset shows that our method achieved a competitive result and was ranked in the top 7 methods on the preliminary test set. Our work is available at: \url{https://github.com/MedCAI/L2SNet}.

\end{abstract}
\section{Introduction}
\label{sec:introduction}
\noindent Fluorodeoxyglucose (FDG) positron emission tomography (PET) combined with computed tomography (CT) is considered the choice imaging modality for diagnosing, staging, and monitoring the treatment responses in various cancers, such as lung cancer and melanoma \cite{townsend2004pet}. This choice arises from PET/CT's ability to leverage the high sensitivity of PET in localizing abnormal tissue function and the specificity of CT in mapping lesions \cite{blodgett2007pet}. Tumor segmentation is an essential step in PET/CT quantitative analysis. However, manual annotation is time-consuming and complicated in the clinic, which limits the effect of PET/CT examination. Fortunately, the development of deep learning gives rise to many automatic tumor segmentation algorithms \cite{huang2022isa,lian2018joint}, which have achieved promising performance in whole-body PET/CT image segmentation.

\begin{figure}[t]
\centering
\includegraphics[width=0.95\textwidth]{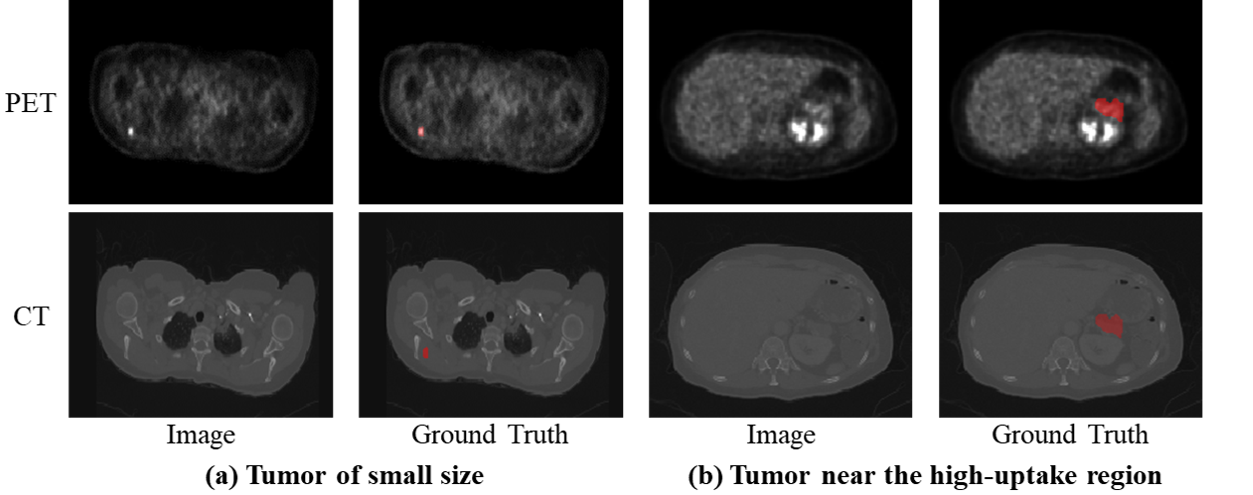}
\caption{Visual illustration of the challenges in automatic tumor segmentation in whole-body PET/CT images. The red region represents the tumor area.} 
\label{fig:challenges}
\end{figure}

Despite the availability of advanced methods, precise tumor segmentation in PET/CT images remains a challenging task, for two major reasons: (i) The small size of tumors poses a difficulty for localizing these abnormal regions, as shown in Fig. \ref{fig:challenges}(a). (ii) As shown in Fig. \ref{fig:challenges}(b), some tumors are adjacent to normal high-uptake regions, causing existing segmentation methods to over-segment the abnormal regions and incorporate areas such as normal high-uptake organs, inflammation, and other infections into the identified tumor regions. Hence, a segmentation method that can accurately localize the tumors and outline their boundaries is urgent for PET/CT image analysis.

To address these issues, we propose a Localization-to-Segmentation framework (L2SNet). Our motivation stems from the fact that, during tumor annotation, clinicians first capture possible lesions in 2D slices and then depict them in the 3D space. Therefore, we argue that localization first and then segmentation is a reasonable solution for distinguishing tumors and normal regions. To achieve this goal, we first predict segmentation results with 2D slices and then use them to guide the 3D segmentation model to focus on the possible tumors. The performance of L2SNet is evaluated in the Automated Lesion Segmentation in Whole-body PET/CT (AutoPET-II) Challenge \cite{autopetii}. 

\begin{figure}[t]
\centering
\includegraphics[width=\textwidth]{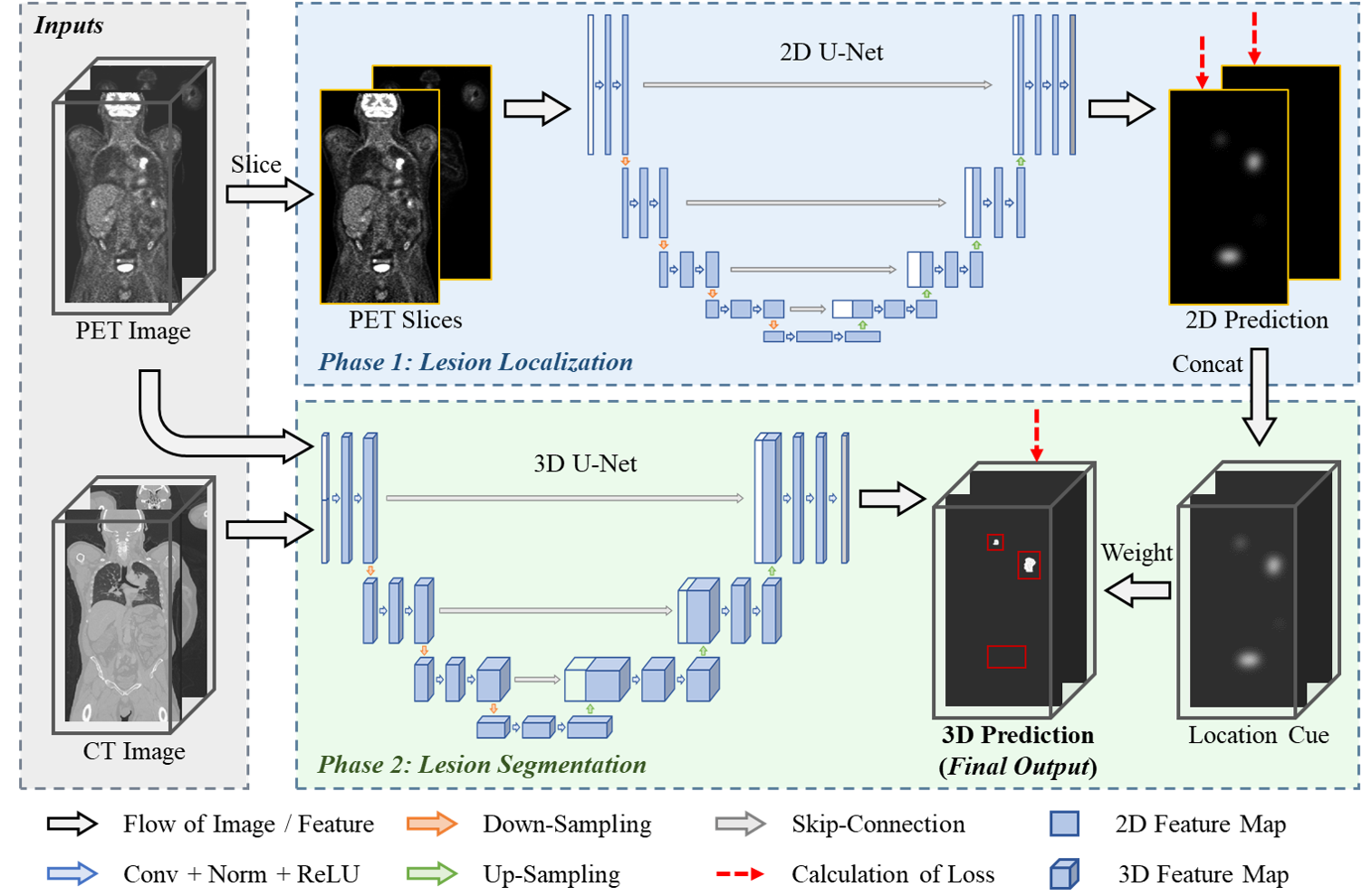}
\caption{Overview of our proposed L2SNet. The red boxes in the 3D prediction reflect the differences between the 3D prediction map and the 2D prediction maps.} 
\label{fig:overview}
\end{figure}

\section{Methods}
\label{sec:method}
\noindent As shown in Fig. \ref{fig:overview}, the proposed L2SNet consists of two phases, namely the lesion localization phase and the lesion segmentation phase. The former takes the 2D PET slices as input and generates the location cue of lesions. Afterward, the lesion segmentation phase uses the location cue as a spatial weight to guide the 3D segmentation network for shaping the lesions. The details of L2SNet are illustrated as follows.

\subsection{Lesion Localization Phase}
\noindent The lesion localization phase produces a coarse segmentation result using coronal slices of PET image. As shown in Fig. \ref{fig:overview}, for an input PET image, L2SNet first splits it into a series of 2D slices. Next, these slices are fed into a 2D U-Net \cite{UNet} for 2D segmentation results, where the feature encoder is an ImageNet-pretrained Res2Net-50 \cite{res2net} and the decoder's channels are set as \{320, 256, 128, 64, 32\} for the $1^{st}$ to $5^{th}$ decoder stages. Each decoder stage consists of two $3\times 3$ convolutional layers with batch normalization and rectified linear unit (ReLU) activation. The up-sampling operation uses the bilinear interpolation to increase the resolution of feature maps. With the final 2D segmentation results, we concatenate them into a location cue to guide the subsequent 3D segmentation.

The design of the lesion location phase considers two problems in existing 3D segmentation frameworks \cite{nnUNet,wang2022psr,zhu2021two}. Firstly, the 3D segmentation methods are powerless to capture the global information of the human body for the large resolution of the whole-body PET image. The coronal slice can reflect the human structure to some extent, and thus the network reduces the false positive rate for the false segmentation on the metabolically active organs such as the brain and kidneys. Secondly, some 3D segmentation solutions use down-sampled 3D Images for training, making the network aware of the global context information. However, this strategy cannot work well for lesion localization in PET images because of the loss of local information and the small size of most lesions. In contrast, the 2D slice strategy does not harm the spatial resolution and thus enables precise lesion location.

\subsection{Lesion Segmentation Phase}
\noindent The lesion segmentation phase uses a 3D U-Net \cite{3DUNet} to make a prediction of the voxel label. Due to the GPU memory limitations, the 3D U-Net is designed to take input patches of size $128\times 192\times 160$ voxels and to use a batch size of two. The 3D U-Net has an encoder and a decoder path, each with four spatial levels. At the beginning of the encoder, patches of size $128\times 192\times 160$ voxels with two channels are extracted from the PET and CT scans as input, followed by an initial $3\times 3\times 3$ 3D convolution layer with 16 filters. The encoder uses convolutional blocks, where each of the blocks consists of two $3\times 3\times 3$ convolutional layers with layer normalization and ReLU activation. The down-sampling uses a max-pooling operation to reduce the resolution of the feature maps by 2. The decoder part has a symmetrical structure with the encoder.

\subsection{Loss Function}
\noindent The training of L2SNet contains two phases. In the first phases, we train the 2D U-Net with binary cross-entropy (BCE) loss function and Dice loss function:
\begin{equation}
\mathcal{L}_{1} = \mathcal{L}_{BCE} + \lambda \mathcal{L}_{Dice} ,
\label{phase1_loss}
\end{equation}
where $\lambda$ balances the two loss functions. In this work, we set $\lambda$ as 1.0 to enhance the ability of the network to capture small lesions.

After the first phase of training, we train the second-stage 3D UNet with BCE loss function and Dice loss function:
\begin{equation}
\mathcal{L}_{2} = \mathcal{L}^{w}_{BCE} + \mathcal{L}^{w}_{Dice} .
\label{phase1_loss}
\end{equation}
Different from the first training loss, we increase the weight of the key voxels according to the corresponding location cue. The weight of the $i^{th}$ predicted voxel can be formulated as $(1 + \mathbf{P}_{i}^{loc})$, where $\mathbf{P}_{i}^{loc} \in [0, 1]$ is the value of the $i^{th}$ voxel of the location cue. Notably, the weights highlight the 3D U-Net's attention to the lesion regions and the false positive regions, which is beneficial for shaping the segmentation boundaries and reducing the false positive rate.

\section{Experiments}
\label{sec:experiment}
\subsection{Data Description}
\noindent Our experiment is based on the AutoPET-II dataset \cite{autopetii}. The dataset contains data from 1014 PET/CT scans examined at the University Hospital Tübingen between 2014 and 2018, where 501 images with malignant lymphoma, melanoma, and non-small cell lung cancer as well as 513 images without PET-positive malignant lesions. We do not use any external data to train our model. 

\subsection{Data Pre-processing and Augmentation}
\noindent L2SNet is composed of two parts, and we have different data pre-processing and augmentation schemes for these two phases. For the lesion localization phase, 3D PET images are sliced into 2D according to the coronal axis, which provides us with a large amount of 2D PET images. However, it raises another noteworthy concern that there are too many slices without tumors. To solve this problem, we do a re-sampling strategy to ensure that there are a certain number of lesion images (about $1/5$) for each training step. Considering the spatial information about the structure of the human body, we do not adopt complex data augmentation. Specifically, we crop the images to $320\times 384$ and then implement random horizontal flip and random rotation in the training phase. In the lesion segmentation phase, we apply intensity scaling for CT (100 to 250) and PET (0 to 15), respectively. Considering the GPU memory limitations, the 3D images are randomly cropped from 243×289×238 voxels to 128×192×160 voxels in patch size. The other 3D data augmentation schemes follow nnUNet \cite{nnUNet}.

\subsection{Training Details}
\noindent Our proposed L2SNet is implemented based on PyTorch 1.13.1 with two NVIDIA\-GeForce RTX 3090 GPUs with 24 GB of memory. As for the maximum number of training iterations, we train the first lesion localization phase for 200 epochs and the second lesion segmentation phase for 500 epochs. We set a batch size of 32 for the 2D U-Net and that of 2 for the 3D U-Net. In both training phases, we use SGD optimizer to optimize the weights, where the learning rate decays in each epoch:
\begin{equation}
\alpha =\alpha_{0} \times (1 - \frac{n}{N})^{0.9},
\label{decay}
\end{equation}
where $\alpha_{0}$ is the initial learning rate of  $1\times 10^{-3}$, $n$ is the current epoch, and $N$ is the total number of epochs. During training, an L2 weight decay of $1\times 10^{-4}$ is applied for regularization.
\subsection{Inference and Post-processing}
\noindent In the inference stage, we use slices for the lesion localization phase and sliding windows for the lesion segmentation phase, generating a location cue and a 3D prediction respectively. To enhance the final performance of the prediction, we adopt a post-processing approach that uses the location cue to adaptively adjust the discrimination threshold for the 3D prediction. Specifically, owing to the sigmoid function, the probability value of each pixel in the prediction maps from both phases varies from 0 to 1. If there is a pixel in the location cue with a probability above 0.5, we will lower the threshold from 0.5 to 0.1 for the 3D prediction map as the pixel is a possible tumor pixel identified by the 2D U-Net and verse vice.

\section{Results}
\label{sec:result}
\begin{figure}[t]
\centering
\includegraphics[width=\textwidth]{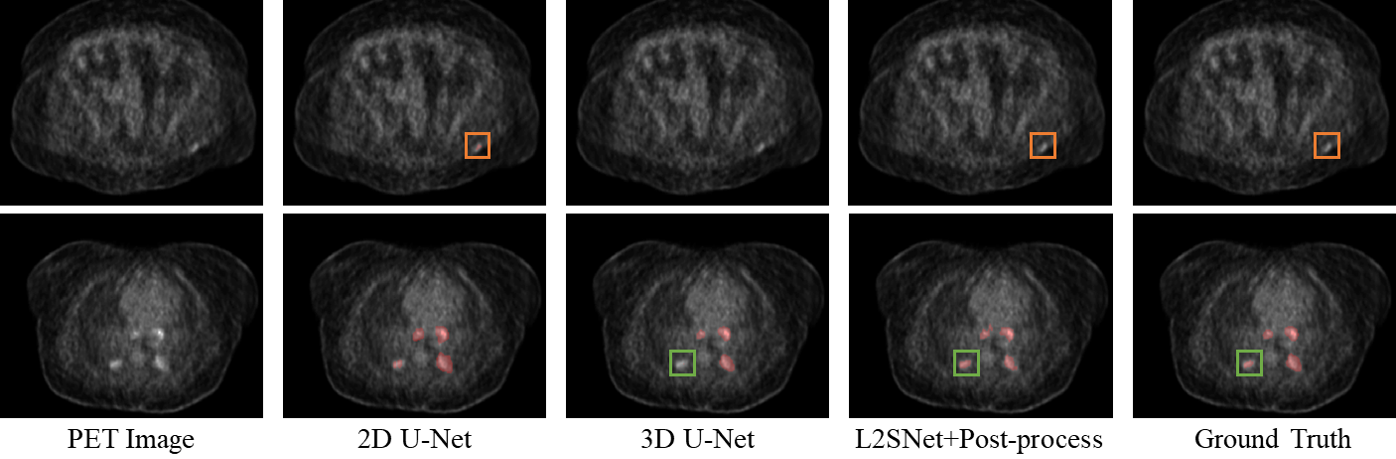}
\caption{Qualitative comparison of different methods. The orange box indicates the false positive region in 2D U-Net result, and the green box indicates the false negative region in the 3D U-Net result.} 
\label{fig:qualitative}
\end{figure}

\noindent We adopt 5-fold cross-validation to evaluate the performance of our method. Following the AutoPET-II challenge \cite{autopetii}, we take mean Dice score (mDice), false positive volume (FPV), and false negative volume (FNV) as evaluation metrics. Table \ref{tab1} lists the quantitative comparison results, from which we can observe that L2SNet achieves superior performance across three metrics compared to the 2D U-Net and 3D U-Net. The post-process strategy further improves the segmentation capability of L2SNet, with an FPV of 1.611 and an FNV of 1.874. Fig. \ref{fig:qualitative} intuitively shows the effectiveness of our method, which takes the segmentation results of 2D U-Net and 3D U-Net into account, generating better segmentation results.

\begin{table}[t]
\centering
\caption{Comparison of tumor segmentation results obtained by 2D U-Net, 3D U-Net, and L2SNet. The discrimination threshold is set to 0.5 for 2D U-Net, 3D U-Net, and L2SNet without post-processing. Best performance is listed in \textbf{bold}.}
\label{tab1}
\resizebox{0.6\textwidth}{!}{%
\begin{tabular}{cccc}
\hline
Methods                  & mDice     & FPV       & FNV                       \\ \hline \hline
2D U-Net \cite{UNet}           & $0.329$  & $21.945$  & $1.803$                   \\
3D U-Net \cite{3DUNet}         & $0.401$   & $3.622$   & $4.204$                   \\
L2SNet   (Ours)                & $\mathbf{0.415}$  & $3.244$   & $3.662$          \\
L2SNet + Post-process (Ours)   & $0.414$  & $\mathbf{1.611}$   & $\mathbf{1.874}$ \\ \hline
\end{tabular}}
\end{table}

\section{Conclusion}
\noindent In this paper, we introduce a localization-to-segmentation framework for the MICCAI 2023 AutoPET-II challenge. Our proposed method consists of two phases, namely, the lesion localization phase and the lesion segmentation phase. In the first phase, we adopt a 2D U-Net for localizing tumors in 2D slices. The second phase uses the location cue to guide the 3D U-Net for shaping the final segmentation result. In the inference stage, we design an adaptive threshold scheme to make full use of the segmentation results of the two phases. Our method achieved competitive results on the preliminary test data. We hope that our method can contribute to the vision community to explore more impressive methods in automatic tumor segmentation of whole-body PET/CT images.

\bibliography{L2SNet.bib}
\end{document}